\documentclass[]{emulateapj}
\usepackage{apjfonts}
\usepackage{natbib}
\usepackage{epsfig}

\newcommand{\lcdm}{\Lambda\mathrm{CDM}}

\newcommand{\gas}{\mathrm{gas}}
\newcommand{\Msun}{M_{\odot}}

\newcommand{\tstar}{t_{\star}}
\newcommand{\Mstars}{\mathcal{M}_{\star}}
\newcommand{\Mstar}{\mathcal{M}}

\newcommand{\Fe}{\mathrm{Fe}}
\newcommand{\aFe}{[$\alpha$/$\Fe$]\,}
\newcommand{\dMdtstar}{\mathrm{d}\Mstars/\mathrm{d}\tstar}
\newcommand{\dMdtstarf}{\frac{\mathrm{d}\Mstar}{\mathrm{d}\tstar}}
\newcommand{\vcirc}{v_{\mathrm{circ}}}
\newcommand{\kms}{\mathrm{km} \, \mathrm{s}^{-1}}
\newcommand{\beq}{\begin{equation}}
\newcommand{\eeq}{\end{equation}}
\newcommand{\beqa}{\begin{eqnarray}}
\newcommand{\eeqa}{\end{eqnarray}}

\shorttitle{$\lcdm$ and Galactic Halo Abundances}
\shortauthors{Robertson et al.}

\begin{document}

\title{
$\Lambda$ Cold Dark Matter,
Stellar Feedback, \\ and the Galactic Halo Abundance Pattern}
\author{Brant Robertson\altaffilmark{1,4},
        James  S. Bullock\altaffilmark{2}, 
	Andreea S. Font \altaffilmark{3},
      Kathryn V. Johnston \altaffilmark{3},
     and Lars Hernquist\altaffilmark{1}}

\altaffiltext{1}{Harvard-Smithsonian Center for Astrophysics, 
        60 Garden St., Cambridge, MA 02138, USA}
\altaffiltext{2}{Department of Physics \& Astronomy,
	University of California, Irvine, CA 92697, USA}
\altaffiltext{3}{Van Vleck Observatory, Wesleyan University, Middletown, CT 06459, USA}
\altaffiltext{4}{brobertson@cfa.harvard.edu}

\begin{abstract}

The hierarchical formation scenario for 
the stellar halo requires the accretion and disruption of
dwarf galaxies, yet low-metallicity 
halo stars are enriched in $\alpha$-elements
compared to similar, low-metallicity stars in
dwarf spheroidal (dSph) galaxies.
We address this primary challenge for the hierarchical formation scenario  for the stellar halo by combining
chemical  evolution  modelling with  cosmologically-motivated  mass
accretion histories for the Milky Way dark halo
and its satellites. 
We demonstrate 
that stellar halo and dwarf galaxy abundance patterns can be explained
naturally  within   the  $\lcdm$   framework.   Our    solution relies
fundamentally on the $\lcdm$ model prediction that the majority of the
stars in the stellar halo were formed within a few relatively massive, $\sim
5 \times 10^{10} \Msun$, dwarf irregular (dIrr) - 
size dark matter halos, which were accreted and
destroyed $\sim 10$ Gyr in the  past.  These systems necessarily have
short-lived, rapid star formation histories, are enriched primarily by
Type II supernovae, and host stars with enhanced \aFe abundances.
In contrast, dwarf spheroidal galaxies exist within low-mass
dark matter hosts of $\sim 10^{9} \Msun$, where supernovae winds are 
important in setting the intermediate \aFe ratios observed.
Our model includes  enrichment from Type  Ia and Type II supernovae as
well as stellar  winds, and includes a physically-motivated supernovae
feedback prescription calibrated to reproduce  the local dwarf  galaxy
stellar mass - metallicity relation. We use  representative
examples of the type of dark matter halos we expect to host
a destroyed ``stellar halo progenitor'' dwarf, a surviving 
dIrr, and a surviving dSph galaxy, and show
that their derived abundance patterns, stellar masses, and gas masses
are consistent with  those observed for each type of system.
 Our model also self-consistently reproduces
the observed stellar mass - $v_{\mathrm{circ}}$ relation for
local group satellites and produces the correct cumulative
 mass for the Milky Way
stellar halo.  We predict that the lowest metallicity stars
in intermediate-mass dIrr galaxies like the SMC and LMC should
follow abundance patterns similar to that observed in
the stellar halo.  Searches for accreted, 
disrupted, low-mass dwarfs may be enhanced
by searching for unbound 
stars with dSph-like chemical abundance patterns.

\end{abstract}


\section{Introduction}
\label{sec:intro}

While evidence in support of a $\lcdm$ universe continues to mount
from high-precision, large-scale cosmological observations
\citep[e.g.][]{spergel03a, blakeslee03a, pen03a, lahav02a, zehavi02a},
a variety of data on galactic and subgalactic scales has remained more
difficult  to explain  \citep[e.g.][and references therein]{simon05a,
moore99a,  klypin99b, navarro97a}.  Among  the biggest  questions in
cosmology is whether  $\lcdm$ needs to    be  modified
substantially   in  order   to  account  for   galactic-scale data
\citep[e.g.][]{spergel00a, kaplinghat00a, colin00a}   or  if  
physical and dynamical processes can
explain   the  data  in  the     context  of the  standard   paradigm
\citep[e.g.][]{bullock00a, alam02a, robertson04a}.
The resolution of this question 
requires testing our cosmological model against the full realm
of existing subgalactic data.
In this paper, we confront
the rich data set of elemental abundances 
in galactic halo and Local Group dwarf galaxy stars
with models set within the $\lcdm$ cosmology.  
We conclude that the
chemical abundance pattern in dwarf irregulars, dwarf spheroidals, and
the Milky Way stellar halo can be explained naturally in the context of
our hierarchical cosmology.

One of the fundamental tenets of the  CDM paradigm is that dark matter
halos form hierarchically, via a series of mergers with smaller halos.
This prediction gives rise to the natural expectation
that the stellar  halo is formed from disrupted,
accreted dwarf galaxies
\citep[][]{johnston96a,helmi99a,bullock01a,bullock05a,font05a}.
Indeed, this picture is remarkably similar to the ideas of
\cite{searle78a} who argued for ``chaotic accretion'' based purely
on observational grounds.  The discovery  of fossil evidence of  these
accretion events via  the  identification of substructure   within the
stellar halo of our Galaxy may provide the  only direct evidence that
structure      formation     is     hierarchical   on     small-scales
\citep{bullock05a}.  Searches for   this substructure  are   under way
\citep[e.g.][]{newberg02a,morrison03a,kinman04a, brown04a, majewski04a}.

One of the main problems facing the hierarchical accretion 
picture for
the stellar halo
 is that abundance patterns
of  surviving local group galaxies do  not match those   of stars in the
stellar halo \citep[see e.g.][\S \ref{sec:abundances}]{wyse04a,venn04a}.  
Stars  in the stellar halo
tend  to   be  old, metal-poor,  and   $\alpha$-enriched,
with \aFe$\sim +0.3$.   
The stellar halo abundance pattern 
contrasts  with the younger  stellar  populations of local, massive dwarf
irregular  (dIrr) galaxies like the  Large  Magellanic Cloud (LMC) and
Small Magellanic  Cloud (SMC), which  are more  metal
enriched and have near-solar [$\alpha$/Fe] ratios
\citep{jasniewicz94a,hill95a,hill97a,hill97b,venn99a,cole00a,hill00a,korn02a,smith02a}.  
Additionally, stars in dwarf spheroidal (dSph) galaxies
around  the  Milky Way
have intermediate  $\alpha$-element abundances
at low-metallicity, \aFe$\sim + 0.15$
\citep{smecker-hane99a,shetrone01a,smecker-hane03a,shetrone03a,venn04a,geisler05a}.
The discrepancy between the ages and metal  abundances of dwarfs and
the stellar halo is often cited as evidence against the
hierarchical 
$\lcdm$  prediction 
\citep[e.g.][]{unavane96a,gilmore98a,fulbright02a,tolstoy03a,venn04a},
and has even been used to argue for a revision of the $\lcdm$ cosmology
\citep{wyse04a}.
 The
problem is clear:  how can one build  the stellar  halo from destroyed
dwarf  galaxies   when the stars  in  low-mass  galaxies appear to be
chemically distinct from the stars in the halo?

Our solution to this conundrum relies fundamentally on two predictions
that  follow  directly from  the  $\lcdm$  cosmology: 1)  most of  the
{\em{mass}} in the Milky Way halo was acquired via mergers with massive
($\sim  5 \times 10^{10} \Msun$) ``dIrr-type''  dark matter halos; and
2) most of these events occurred quite early  on, at a look-back time of
$\sim 10$ Gyr.  It follows naturally that the majority of stars in the
stellar   halo were formed  in  intermediate-mass  halos  of this type
\citep{bullock05a}.
The associated galaxies were disrupted soon after accretion and before significant chemical evolution.

Within the context of our scenario, the
observed tendency for halo stars to be metal poor
and $\alpha$-enhanced compared  to  stars in
 surviving dIrr  galaxies  follows
from standard  assumptions   about  Type Ia  vs.    Type II
supernovae  enrichment.  Type II supernovae provide $\alpha-$enhanced
enrichment    within   $\sim 10$ Myr  of    a
star-formation episode  (see \S \ref{sec:yields}), while the iron-rich
nucleosynthetic products  from  Type  Ia  supernovae   contribute much
later, at $\sim 1$Gyr.  The massive, ``stellar halo progenitor'' halos
are accreted early and destroyed soon after.  These systems
are forced to have short, truncated star formation histories,  and   
naturally  host stars     that  are predominantly
Type-II enriched and $\alpha$-enhanced.
In contrast, surviving dIrr
galaxies are accreted much later than their
destroyed counterparts (this is what allows them to
survive, see \S \ref{sec:cosmo}), 
form stars for many billions  of years, and 
have substantial enrichment from both Type Ia and Type II supernovae.
Finally,
lower-mass,  dwarf spheroidal galaxies inhabit  halos with 
much shallower potential wells  ($\lesssim 10^{9}  \Msun$) than the
massive dIrr galaxies.  We  argue  below
that blow-out   feedback    from supernova give   rise  to   both  the
low-metallicities    observed   in these systems    as  well  as their
intermediate [$\alpha$/Fe] ratios.

In  what follows, we develop a  model to track the chemical enrichment
histories in dwarf  galaxies forming in cosmologically self-consistent
dark matter halos.  We use   three example mass accretion   histories.
Two are chosen to  be representative of the  kind of dark matter halos
we    expect  to host  {\em{surviving}} 
dSph and  dIrr  satellites, based on $\lcdm$
predictions.  A third example accretion history
is typical of the type that is accreted and  destroyed early 
to contribute to the formation of the halo.   We
track chemical enrichment by including  contributions from Type Ia and
Type II supernovae  as well as stellar winds.   Our star formation and
feedback  model self-consistently  reproduces: 1) the  total   mass of the
stellar  halo \citep{bullock05a}; 2)  the stellar mass-$v_{\mathrm{circ}}$ relation
expected for Local  Group dwarfs; and 3) the mass - metallicity relation
found by \cite{dekel03a} (see \S \ref{sec:gce}).
We go on to use this model to interpret the chemical
abundance patterns of the  stellar halo, dwarf spheroidals,   and dwarf
irregular galaxies
\citep{venn04a}.

We note that there exists an extensive literature of galactic chemical
evolution modelling
\citep{van_den_bergh62a,larson72a,tinsley72a,talbot73a,pagel75a,hartwick76a,tinsley80a,twarog80a,matteucci86a,matteucci87a,matteucci89a,wheeler89a,ferrini92a,timmes95a,tsujimoto95a,devriendt99a,ferreras00a,chiappini03a,lanfranchi04a}.
We aim neither to critique nor supersede this previous work.  Our goal
is simply to approach galactic chemical evolution with  a hierarchically
driven and cosmologically  consistent method while  retaining complexity
in our chemical enrichment code \citep[see also][]{brook04a,brook04b}.  By placing the chemical enrichment
of dwarf galaxies  into the  context  of their expected   cosmological
formation and subsequent fate, we hope to gain some insight on
interpreting   the rich data  sets   being compiled  from Local  Group
observations.

In the next section, we provide  an overview of the expected formation
histories  of Milky  Way-type dark  matter   halos, the accretion  and
disruption histories of their  accreted subhalos, and the mass
accretion histories  of    the satellite  halos    themselves.   In \S
\ref{sec:examples} we discuss our  star formation law and  present our
three example mass accretion histories used to facilitate our comparisons to dIrr, dSph, and stellar halo abundance data.
We detail our stellar yield modelling in \S \ref{sec:yields},   
and our  chemical    evolution  and feedback calibration in 
\S \ref{sec:gce}.
 We present our results 
in \S \ref{sec:abundances}  and summarize and  conclude
in \S \ref{sec:conclusions}.  Throughout this paper we assume a flat
$\lcdm$ cosmology with $\Omega_m = 0.3$, $\Omega_{\Lambda} = 0.7$, 
$h = 0.7$, and $\sigma_8 = 0.9$.

\section{Cosmological Context}
\label{sec:cosmo}

We  first  review several relevant expectations   for the formation of
galaxy-size dark matter halos within  the $\Lambda$CDM framework.  The
reader is referred to \cite{wechsler02a, zentner03a} and
\cite{zentner05a}
for more detailed discussions.

\subsection{Hierarchical growth of the Milky Way dark halo}

Consider a galaxy-size dark matter halo of mass  $M_{\rm h}$ at $z=0$.
Mass growth in halos of this type proceeds  by the merging and accretion of
a  wide mass   spectrum   of  ``subhalos''  or   ``satellite  halos''.
Integrated over the  lifetime of the host  halo, the accreted spectrum
of  satellite halo masses, $M_s$, rises softly towards the low-mass end,
 $d  N / dM_{\rm s} \propto M_{\rm s}^{-1/2}$
and falls off sharply above a characteristic cutoff mass
$\sim 0.1 M_{\rm h}$.  Owing to the shallow slope,
most of the {\em mass} accreted
onto the host is associated with subhalos of roughly the
cutoff mass. 

The Milky Way's dark matter halo
 has  a virial mass of  $M_{\rm h} \simeq 10^{12}
\Msun$ at $z=0$  and is thus expected to  have  accreted most of its  
mass via the accretion of intermediate size systems $M_{\rm s}
\simeq  (1-10) \times 10^{10}  \Msun$ \citep{zentner03a}.
Accreted subhalos lose   
mass as a  result of  violent  interactions with the host halo
system, and are typically destroyed within a few dynamical times.
The stripped material incorporates into and effectively builds the
background dark matter  halo of the  host.  
The model of \cite{bullock05a} that serves as
our cosmological framework accounts for these dynamical effects, in
addition to the redshift evolution of the characteristic mass and densities
of merging systems.

The time line of accretion activity is sensitive to the host halo mass
and the  governing cosmological parameters.   For Milky-Way size halos
in the standard $\Lambda$CDM model, the accretion rate of dwarf-galaxy
halos peaks $\sim 10$ Gyr in the past,  and slows down considerably at
a look-back time of $\sim 8$ Gyr, although there is scatter
from halo to halo \citep{zentner03a}.  Satellite halos that get
accreted during  the earliest epochs have  time to  be destroyed  by
interactions with the host galaxy's  environment
\citep[see][]{zentner05a}.
The increased efficiency of dynamical friction with satellite mass causes more massive subhalos to lose mass
more quickly than smaller subhalos.  
Most of  the subhalos
that  survive until   the  present  day   were  accreted   during  the
slow-growth  phase, within   the  past  $\sim  5$ Gyr,  and  therefore
represent a temporally biased  population compared to  their destroyed
counterparts.
We  emphasize that this temporal  bias between surviving and
destroyed systems implies that the  two populations have intrinsically
different  formation histories: the  destroyed population of halos are
expected to  have formed more rapidly  than  the
subhalos that survive.

We  expect  that   the  most  massive,  accreted
satellite halos  will host dwarf galaxies, and  that most of these
systems  will be  destroyed by  tidal   interactions.  
Just as the dissipationless dark
halo of the host is  built from the dark matter that was originally
associated with accreted and destroyed subhalos,
the stellar halo is built from the
dissipationless {\em stellar} matter that was associated
with the same population of destroyed galaxies \citep[][]{bullock01a}.
Given this scenario for stellar halo formation,
two important implications follow
directly from the $\Lambda$CDM predictions discussed above:
\begin{enumerate}
\item{The majority of the stars 
contained in the Milky Way stellar  halo 
originated in $\sim 5 \times 10^{10} \Msun$ ``dwarf irregular-type''
dark matter halos.}
\item{The surviving satellite galaxies around the Milky Way
represent a {\em biased} population:
they typically were accreted later than the dwarf galaxies that
were destroyed to make the halo.}
\end{enumerate}
The second point
implies that the mass accretion and star formation histories
of surviving satellites
are more gradual than their destroyed counterparts.

\subsection{Dark matter halo mass accretion histories}
\cite{wechsler02a} used N-body
simulations to study the mass accretion histories of  dark matter
halos observed at various redshifts.    One of the main results
of their work was to show  that
the cumulative mass accretion in individual dark  matter halos  
can be well-described by a simple function:
\begin{equation}
M(a) = M_0 \exp \left[ -2 a_c \left(\frac{a_0}{a} - 1 \right) \right],
\label{eqt:mofa}
\end{equation}
where $a = (1+z)^{-1}$ is the expansion factor and
$M_0$ is the mass of the halo as observed at
the epoch $a_0  = (1+z_0)^{-1}$.  
Given a halo mass and redshift, this function has only
one parameter, $a_c$,  which is a characteristic  formation 
epoch for the halo.   The
typical  value of $a_c$  varies as a function of  halo mass, with more
massive  halos having  earlier formation times.
At a  fixed $M_0$, the  distribution of $a_c$ is roughly
log-normal, with a scatter of $\Delta \log_{10} a_c = 0.14$.
Of note is the interesting fact that the  
$a_c$ distribution is independent of $a_0$ for halos of a fixed mass.

Wechsler et al. (2002) showed that
the median value of the formation epoch parameter, $a_c$,
can be determined for a halo of mass $M_0$ using a simple model
based on spherical collapse.
Specifically,  $a_c$ is set
by the epoch at which $M_{*}$, the typical collapsing mass
\citep[e.g.][]{lacey93a}, 
equals a fixed fraction $F=0.015$ of the halo mass $M_0$:
\begin{equation}
M_{*} (a_c) \equiv F M_0.
\end{equation} 
As expected, this expression for $a_c$ is independent of $a_0$.
A Milky-Way type halo, for example, has a median $a_c \simeq 0.30$,
corresponding to a look-back time of $\sim 10.5$ Gyr, which is consistent
with the expected epoch of peak accretion discussed above.

\section{Star formation and example histories}
\label{sec:examples}

We will use three example mass accretion histories to explore
expected differences between halo stars and surviving satellites.
Since each system under consideration is a low-mass object with 
efficient gas cooling, we assume the cold gas inflow rate, $g(t)$,
tracks the dark matter accretion rate:
\beq
g(t) = f_{\gas} \, h(t - t_{\rm in}).
\eeq
Here, $f_{\gas}$ is  the fraction of the  dark matter halo mass in the
form of cold, accreting  gas, and $h(t) =  {\mathrm d} M/{\mathrm d}t$
is  the  halo  growth  rate determined  by  taking  the derivative of
Equation  (1) with  respect  to cosmic time.   The  time  lag $t-t_{\rm in}$ between
$h(t)$ and  $g_(t)$ accounts for the finite  time it takes for gas to
settle into the central galaxy after being accreted onto the halo.  We
assume this settling occurs  in roughly a dynamical time
at  the
virial radius: $t_{\rm  in} = R_{\rm  h}/\vcirc  \simeq 2 \,  {\rm
Gyr} \,  \, (1+z)^{-3/2}$.  We set  $f_{\gas} =  0.02$, in accord with
the   cold  baryonic mass fraction    in  observed galaxies  \citep{bell03a}.

The star formation rate, $\psi(t)$, in each system is set using
the simple relation
\beq
\psi(t) = \frac{M_{\gas}}{\tau_{\star}},
\label{eq:sfr}
\eeq
where $M_{\gas}$ is the total cold gas mass in the system and
the star formation timescale, $\tau_{\star}$, is a free parameter.
In the models we explore below we assume for simplicity that star
formation  is   truncated soon after  each satellite halo
is accreted onto the Milky Way host  as a  result   of
ram-pressure stripping from the    background hot gas halo
\citep{moore94a,maller04a}.  Of  course,  this assumption is
oversimplified and cannot be accurate in detail,  but it allows us to
capture the general differences we expect from one example to the next
without addressing the complicated interplay between orbital evolution
and  star formation.
Ram-pressure will be more effective
at stripping gas from low-mass objects with shallow potential wells
(dSph) while realistically dIrr-size halos will tend to retain their
gas more efficiently.
Additionally, for a given mass, typical encounter velocities are
higher at higher redshifts.  Surviving dwarf satellites will undergo
correspondingly less ram-pressure stripping than their counterparts
that are accreted into the halo early and subsequently disrupted.
However, for massive systems that were accreted early (a ``halo-progenitor'', for instance) our assumption of star formation truncation will not
qualitatively affect our results compared to a model where star formation
is allowed to continue until the galaxy is tidally destroyed.  Early-accreted
systems are typically destroyed very soon after they merge (over a $\sim$ 1 Gyr timescale, a typical dynamical time for a z$\sim$1 halo).
Nevertheless, we acknowledge our assumption of truncated star formation
is simplified.  We aim to explore the implications of varying models of
star formation within satellite halos in subsequent work.

The gas  mass  in Equation
\ref{eq:sfr} will not simply be the cumulative 
integral of gas  inflow rate, $g(t)$, as we allow for 
the  effect of  star  formation feedback and associated gas
loss within each satellite  halo.
The system of  equations we use  to
track star  formation, gas loss, and  metal enrichment is presented in
\S 4.    Including this gas outflow,
we find that  the choice $\tau_{\star}  = 6.75$ Gyr  allows us to match
the observed  mass-to-light ratios  in  dIrr-sized objects,  and, when
incorporated into our full mass accretion  history models of the Milky
Way, the same choice gives a good match to the total Milky Way stellar
halo  mass as well  as its radial profile  \citep{bullock05a}.
The inclusion of feedback in our model helps regulate the star formation
rate in dwarf galaxies.  As a result, we must adopt a smaller star formation
timescale than that required in the similar model utilized by \cite{bullock05a}
which does not explicitly use a prescription for feedback from star formation.
Before moving   on to  discuss our  treatment   of outflow   and metal
enrichment   in detail,  we first introduce   our  three example  mass
accretion histories.

\bigskip

\noindent {\bf  Dwarf Irregular Satellite} \\
\noindent 
For our surviving, ``dIrr'' satellite halo  example, we use a dark
matter  halo of  virial mass of  $M_0 = 6 \times   10^{10} \Msun$, which  we
assume is accreted into  the Milky Way halo at  $a_0 = 0.79$. 
This accretion epoch corresponds  to a  look-back time  of $3.1$  Gyr,
a typical  accretion  time for
surviving, massive  satellites (A. Zenter \& J. Bullock, private communication)
 and close to that
expected for the LMC from  gravitational and hydrodynamical
simulations of its orbital evolution within the Milky Way
potential \citep{mastropietro04a}.
 A halo  of this mass at $a_0  = 0.79$ has a
median maximum circular velocity of $\vcirc = 72  \, \kms$ \citep{bullock01a},
 but we  expect that dark  matter  mass loss will leave  this
system with a   circular velocity closer to  that  of the SMC  or  LMC
\citep[$\sim 60 \kms$;][]{van_der_marel02a} after three billion years
of evolution \citep[][]{bullock05a}.

We  assume
that the dark matter mass accretion  history of our dIrr halo follows
the form of Equation 1, with a  characteristic formation epoch of $a_c
=  0.231$  (Equation 2),  corresponding  to  a  look-back time  of
$11.6$ Gyr.  Note that the time
between ``formation'' at $a_c$ and  its accretion into the
Milky  Way dark halo at  $a_0$, this system  will have had
$\sim 8.5$ Gyr to form stars -- ample time for both Type Ia and
Type II enrichment.
 After applying the star formation law 
discussed above and the feedback prescription described in \S 4, the final stellar mass
and gas mass in this system are $M_{\star} = 3.9 \times 10^{8} \Msun$ and
$M_{\gas} = 4.7 \times 10^{8} \Msun$,
respectively.  Allowing for a small amount of gas loss
from ram pressure stripping after accretion
\citep[see e.g.][]{mastropietro04a},
these numbers
are in line with gas mass and stellar mass measurements
for the  SMC and LMC
 \citep[e.g.][]{grebel01a, grebel03a}.
\\

\noindent {\bf  Dwarf Spheroidal Satellite}\\
\noindent
Our Dwarf Spheroidal example consists of a relatively low-mass halo,
$M_0 = 5.6 \times 10^8 \Msun$, accreted at $a_0 = 0.667$ or at
a look-back time of $5$ Gyr.  
The accretion  time is typical for that of
surviving satellite halos of this size (A. Zentner \& J. Bullock, private communication).
 The halo's formation epoch is
$a_c = 0.08$, corresponding to a look-back time of $13.1$ Gyr.
This system
has a maximum circular velocity of $\vcirc = 20 \kms$ at accretion,
which will likely be reduced to $\vcirc \simeq 15 \kms$ after
five billion years of evolution within the galaxy potential
\citep{bullock05a}.
This circular velocity agrees with circular velocity
estimates for dSph satellite galaxies in the Local Group
\citep[e.g.][]{zentner03a}.  The time available for star
formation between $a_c$ and $a_0$ is $\sim 8$ Gyr for this system.

  It is now well-known that
if $\Lambda$CDM is   correct, only roughly one-in-ten   of the
dwarf-size satellite
halos   can host a dwarf  galaxy, with the
rest remaining dark \citep{moore99a, klypin99b}.
A natural solution to this
problem relies on the fact that halos
of this size ($\vcirc \lesssim 30 \kms$) are unable
to accrete gas after the universe becomes reionized
\cite[e.g.][]{bullock00a}.  We account for this effect
by setting the gas accretion rate
$g(t)  = 0.0$ after  the epoch of reionization, $a_{\rm  re}$. 
The value of  $a_c$ we  have adopted for
our dSph example  is $\sim 2  \sigma$ earlier than the median expected
for a halo of its size (via Equation 2).  We have  made this choice in
order to ensure  that this halo  collapses  before cosmic reionization
and is able to host a visible dwarf galaxy.
Our assumption about the effect of reionization 
on dwarf galaxy formation allows
our  picture  to self-consistently  address  the ``dwarf  satellite
problem'' without modifying the tenets of $\Lambda$CDM.
For
concreteness,  we choose  $a_{\rm re}  =  0.09$,  but our  results are
insensitive to  this choice.  

The stellar mass and gas mass within this system at accretion is
$M_{\star} = 1.1 \times 10^6 \Msun$ and $M_{\gas} = 3.8 \times 10^5 \Msun$
respectively.
The stellar mass is consistent with that of dSph galaxies in the local group,
and its gas mass fraction is lower than that in the dIrr example, also in accordance with observed trends \citep[e.g.][and references therein]{mateo98a, grebel03a}.  
We expect that the gas mass in this system will be further reduced
after accretion as a result of efficient ram pressure stripping
from the shallow host halo potential well.
\\

\noindent {\bf  Stellar Halo Progenitor}\\
\noindent
As argued in \S 2.1, and discussed in more detail in
\cite{bullock05a}, the hierarchical scenario
predicts that most of the mass in the stellar
halo was contributed by several massive subhalos that were accreted
onto the Milky Way host long ago.
We choose to  model one such subhalo as our example ``stellar halo progenitor'' a dark halo with
a virial mass equal to that of our dIrr example above, $M_0 = 6 \times
10^{10} \Msun$, but which was  accreted $9$ Gyr in  the past, at $a_0 =
0.417$.  Because this subhalo has the same mass as the dIrr example, 
the stellar halo progenitor has
the same formation epoch,
$a_c = 0.231$.  However, because the subhalo was accreted (and destroyed)
at a much earlier time, 
the time available for star formation in this system 
is only $\sim 2.6$  Gyr, and we therefore 
expect that the $\sim 1$ Gyr required for Type Ia enrichment
will limit significant Type Ia enrichment in this system.

When we apply our star formation and feedback model,
the stellar mass at accretion in this system is
$M_{\star} = 2.4 \times 10^8 \Msun$ and the cold gas mass 
is $M_{\gas} = 7.2 \times 10^8 \Msun$.  The high gas fraction is
a result of have a shorter time available to turn gas into stars.
Indeed, since a majority of the baryons in the Milky Way system
were accreted in early events of this size (\S 2.1), 
gas-rich mergers of this type are likely fundamental in fueling
the growing Milky Way bulge and disk components.
Although a large number of lower-mass subhalo accretion events will likely contribute significant substructure to the Galactic halo, we expect that 
$\sim 4-5$ mergers of subhalos similar to our halo progenitor model are responsible
for most of the {\em{mass}} of the Milky Way stellar halo.

\bigskip

\begin{figure}
\figurenum{1}
\epsscale{1}
\plotone{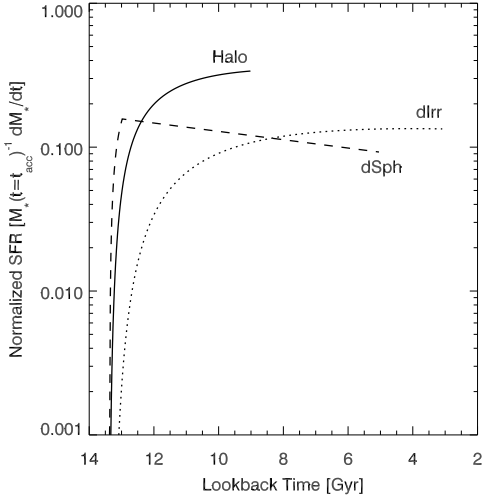}
\caption{\label{fig:sfr} 
Model star formation histories of our halo-progenitor example
(solid line), dIrr  (dotted line), and low-mass dSph galaxy (dashed line).
The halo-progenitor model forms stars quickly until it is
accreted  into the stellar  halo $\sim 9$ Gyr  ago.  The surviving
dIrr galaxy forms stars more  gradually, until its
recent accretion into the galaxy $\sim 3$ Gyr ago.
The   dwarf spheroidal   galaxy has   its  gas  infall  suppressed  by
reionization early-on and converts  its gas mass  into stars 
until its accretion into the galaxy $\sim 5$ Gyr ago. Both the dIrr and
dSph are expected to survive as Milky Way satellites until the present day.
}
\end{figure}

In Figure  \ref{fig:sfr}, we plot 
star  formation histories as a function of look-back time
for each example history: halo-progenitor (solid  line), 
dwarf irregular (dotted  line),   and 
  dwarf  spheroidal
(dashed line).  The break in the dSph track at
$\sim 13$ Gyr is associated with suppression of gas infall at
reionization. 
From  Figure \ref{fig:sfr}   we   see that  the  halo-progenitor  star
formation  history is truncated and more  rapid compared with the
surviving  dIrr.  The  history  of  the  dSph example  is intermediate
between the two.  We now address how these differences 
will manifest themselves as chemical signatures in the different
stellar systems.

\section{Stellar Yields}
\label{sec:yields}
In order  to  track   the  time-dependent chemical content   of  dwarf
galaxies during their  formation,  we model  the  chemical yields from
stellar populations, including Type Ia  supernovae, Type II supernovae,
and    stellar winds  from intermediate   mass  stars.    The  rate of
enrichment is  calculated analytically and   the yields for individual
chemicals are selected from the literature.

\subsection{Initial Mass Function}
We define our initial mass function (IMF) as
\begin{equation}
\phi(m) \equiv \frac{\mathrm{d}N}{\mathrm{d}m} = B(m/\Msun)^{-s},
\end{equation}
\noindent
which provides the differential number of stars of mass $m$.
We adopt the \cite{kroupa01a,kroupa02a} IMF, which has a
mass-dependent slope $s$ given by
\begin{equation}
s  = \left\{ \begin{array} {r@{\quad:\quad}l}
	0.3 & 0.01 \leq m < 0.08 \\ 1.3 & 0.08 \leq m < 0.5 \\  2.3 & m \geq 0.5 \end{array} \right.
\end{equation}
\noindent
We set the constant $B$ to normalize the IMF to give
\begin{equation}
\int_{0.1}^{100} m \phi(m)\mathrm{d}m = 1 \Msun,
\end{equation}
\noindent
as we consider stars of mass $0.1\leq m \leq 100$.
Our normalization   provides $\phi(m)$ in   units  of $\Msun^{-1}$ and
allows us to calculate our yields per unit solar mass of stars.
We note that for stellar masses $m > 0.5 \Msun$ the Kroupa IMF is 
very similar to the \cite{salpeter55a} IMF that has a constant
slope of $s = 2.35$, though, as discussed by \cite{kroupa02a}, uncertainties
in determining the IMF of massive stars may mask a steeper slope
of $s \approx 2.7$ for stars above 1 $\Msun$.

\subsection{Type II Supernovae}
\label{sec:yields:snII}

Given an IMF,
we  can track the cumulative number  of Type II supernovae
that have occurred per solar mass of star formation 
as a function of time $t$ as
\begin{equation}
N_{II}(<t)     =                    \int\limits_{\Mstars(t)           
> \Mstar_{II,\mathrm{min}}}^{\Mstar_{II,\mathrm{max}}}\phi(m) \mathrm{d}m.
\end{equation}
\noindent
Here, $\Mstars(t)$ is the mass of a star with  stellar  lifetime $t$
and $\Mstar_{II,\mathrm{min}}$ is the minimum mass star producing a
Type II supernova and $\Mstar_{II, \mathrm{max}}$ is the maximum stellar
mass that will produce Type II yields.  We assume that all stars
more massive than $\Mstar_{II, \mathrm{max}} = 50 \Msun$ will collapse
completely into black holes and we use $\Mstar_{II,\mathrm{min}} = 10 \Msun$.

For the stellar lifetime as a function of star mass, we adopt
a  dual power-law fit  to the \cite{schaller92a} stellar lifetimes
of the form
\begin{equation}
\tstar(\Mstar) = 10.5\left(\frac{\Mstar}{M_{\sun}}\right)^{-3} + 0.035\left(\frac{\Mstar}{M_{\sun}}\right)^{-0.55}
\end{equation}
with $\tstar$ in Gyr.  The the inverse quantity $\Mstars(t)$ and its
derivative $\dMdtstar$  can be obtained   from this relation.   
The  rate  of Type II   supernovae  per solar mass  in  each
stellar population is
\begin{equation}
R_{II}(t) = -\dMdtstarf \phi [\Mstars(t)],
\end{equation}
\noindent
with  no further  normalization  being necessary  as all massive stars
between $\Mstar_{II,\mathrm{min}}$ and 
$\Mstar_{II,\mathrm{max}}$ are considered
to produce supernovae.  

The instantaneous chemical yield of an element $X$ from Type
II supernovae as  a function of time after  a stellar population forms
will   then  follow 
\beq
Y_{_{X,II}}(t)   =   y_{_{X,II}}\left[\Mstar(t) \right] \, \, R_{II}(t),
\eeq
where $y_{_{X,II}}[\Mstar]$ 
is the Type II yield of element $X$ from  a star of mass
$\Mstar$.  In practice, we use the Type  II yields from \cite{tsujimoto95a}
and \cite{thielemann96a}, interpolating between their simulated models 
in order to obtain yields as a continuous function of mass.

Given the above model,
Type II  supernovae  provide enrichment from $\tstar(50)=4.1$  Myr
until $\tstar(10)=20.4$ Myr after each stellar population forms.

\subsection{Type Ia Supernovae}
\label{sec:yields:snI}
In calculating the rate of Type Ia supernova, we follow the formalism
developed by \cite{greggio83a}, which postulates 
that the progenitors of
Type Ia  supernovae  are  binary star systems.  
The   rate of  Type  Ia
supernovae  is  set by  the evolution of   the  secondary star in each
binary system.

We define the number of  Type  Ia supernovae that have
occurred before time $t$ as
\begin{equation}
N_{I}(<t)                            =                               A
\int\limits_{\Mstar_{b,\mathrm{inf}}}^{\Mstar_{b,\mathrm{max}}}\phi(\Mstar_{b})\left[\int\limits_{\mu_{\mathrm{inf}}}^{1/2}
2^{1+\gamma}(1+\gamma)\mu^{\gamma}\mathrm{d}\mu\right] \mathrm{d}\Mstar_{b},
\end{equation}
\noindent
where $A$ is a  normalization parameter.  The outer integral sums
over the total binary mass, 
$\Mstar_b = \Mstar_1 + \Mstar_2$,with $\Mstar_2 < \Mstar_1$, and
the term  in brackets describes   the
distribution of secondary  mass  fractions 
$\mu = \Mstar_{2}/\Mstar_{b}$.  The
exponent $\gamma=2$ sets  the
spectrum   of secondary masses.
The upper limit on the total binary mass is set 
by twice
the maximum primary mass that allows the development of
a degenerate C/O core before filling its Roche lobe,
 $\Mstar_{b,\mathrm{max}}=16 \Msun$.
The lower limit forces
$M_2 \le M_1$ via
$\Mstar_{b,\mathrm{inf}}=\mathrm{MAX}[\Mstar_2(t),\Mstar_{b,min}]$
with $\Mstar_{b,\mathrm{min}}=3\Msun$ following
\cite{greggio83a}.
As  in \S 3.2, 
$\Mstar_2(t)$ is the mass corresponding to the
stellar lifetime $t$.
The lower limit on the secondary mass fraction
is $\mu_{\mathrm{inf}}=\mathrm{MAX}[\Mstar_2(t)/M_{b},(M_{b}-\frac{1}{2}M_{b,\mathrm{max}})/M_{b}]$.
In   order   to
approximate local supernovae rates as determined by \cite{tammann82a},
we follow  the \cite{matteucci86a} Type I SN normalization.
Adjusting the \cite{matteucci86a} result for our chosen IMF,
we find $A\approx0.0956$.

Given the cumulative number of Type Ia supernovae as a function of time
$t$, the corresponding rate is simply
\begin{equation}
R_{I}(t) = \frac{\mathrm{d}N_{I}(<t)}{\mathrm{d}t}.
\end{equation}
\noindent
The instantaneous chemical yield  for each element $X$ from  Type Ia supernovae is then
$Y_{_{X,I}}(t) =   y_{_{X,I}}R_{I}(t)$.   For  the   individual   yield  per
supernovae $y_{X,I}$, we  adopt the  yields of the  Nomoto W7  Type Ia
supernova model \citep{nomoto84a,nomoto97a}.  In this model the Type Ia
supernovae provide enrichment  for  $t>\tstar(8)=31.7$ Myr after  each
stellar population forms, with the peak of the Type I rate occurring 
at $t \sim 1$ Gyr.

\subsection{Stellar Winds}
\label{sec:yields:winds}
We follow  the chemical enrichment from  stellar winds by tracking the
death rate of stars and assuming the majority of the stellar mass loss
occurs during asymptotic giant branch thermal pulsing  near the end of
the stellar lifetime.  Our effective rate of stellar winds is then
\begin{equation}
R_{W}(t) = -\dMdtstarf \phi(\Mstar),
\end{equation}
\noindent
which, like the Type II supernovae rate, is set by the IMF and stellar
lifetimes.  The instantaneous chemical yield of an element $X$ from stellar winds is
then    $Y_{_{X,W}}(t)    =   y_{_{X,W}}R_{W}(t)$.      We    utilize    the
\cite{van_den_hoek97a} time-dependent  stellar  wind yields $y_{X,W}$,
interpolating over  mass  and metallicity  for stellar  masses between
$0.8-8\,\Msun$.  Stellar  winds   provide  enrichment in   this   model
continuously after $\tstar(8)=31.7$ Myr.

\section{Chemical Enrichment and Feedback}
\label{sec:gce}

\subsection{Tracking Abundances}
We aim to track the abundances
of H, He, Fe, and the $\alpha$-elements O and Mg using
the three example accretion histories discussed in \S 2.
To do so, we follow the evolution of stellar  mass
$M_{\star}$,   gas mass $M_{\mathrm{gas}}$,  and mass  $M_{_X}$ in 
each element   $X$ using the system of equations:
\begin{eqnarray}
\frac{\mathrm{d}M_{\star}}{\mathrm{d}t} &=& \psi(t) - Y(t) \\
\frac{\mathrm{d}M_{\mathrm{gas}}}{\mathrm{d}t} &=& -\psi(t) + Y(t) +g(t) - e(t)\\
\frac{\mathrm{d}M_{_X}}{\mathrm{d}t} &=& -f_{_X}\psi(t) + Y_{_X}(t) + h_{_X}g(t) - f_{_X}e_{_X}(t).
\end{eqnarray}
\noindent
As discussed in \S 2.3, the star formation rate is assumed to track the
mass in cold gas as $\psi(t) = M_{\gas}/\tau_{\star}$, and
the gas infall rate, $g(t)$, 
tracks the halo mass accretion rate.
The quantity $Y(t)$ is the total rate of mass input into the gas
from supernovae of both types and stellar winds, 
summed over all stellar populations.
Similarly, $Y_{_X}(t)$ is the instantaneous input rate of the specific
element $X$, again summed over supernovae and winds for all stellar
populations.  
The gas mass fraction in each element $X$ is $f_{_X}  =
M_{_X}/M_{gas}$,  and 
$h_{_X}$ is the elemental mass fraction in {\em{newly accreted}} 
gas.   The ejection rate for each element is chosen
to follow the  star formation rate as
$e_{_X}(t) = w_{_X}\psi(t)$, with $w_{_X}$ representing the efficiency of
gas   blow-out  for element  $X$.  The  total  gas  ejection rate in the
second equation is then the sum over all elements $X$, 
$e(t) = \Sigma f_{_X}e_{_X}(t)$. 

Each
star formation history begins by
accreting  low-
metallicity, alpha-enriched gas, with 
[Fe/H]$=-4.0$ and [$\alpha$/Fe]=0.3.
 Throughout the history of each galaxy, we assume
that the infalling gas retains this initial Type
II-enriched abundance pattern 
(this sets $h_{_X}$ in Equation 17), although, in detail,  our
model   results are   insensitive   to the   abundance  pattern of the
low-metallicity infalling gas.
In the results presented in \S \ref{sec:abundances},
we define our  abundances relative  to the
solar abundances measured by   \cite{grevesse98a}.
  We turn now to a more in-depth
discussion of our treatment of blow-out feedback.  Our feedback
prescription is important for explaining the abundance pattern
in low-mass, dSph-type galaxies.

\subsection{Feedback and the Mass-Metallicity \\ and Mass-$v_{\mathrm{circ}}$ Relations}
\label{sec:dw}
Gas  and metal   ejection from supernovae winds   almost
certainly  plays an  important role in  
shaping the properties of galaxies -- especially dwarf-sized galaxies
\citep{dekel86a}. 
\cite{dekel03a} (DW) have  used a compilation of dwarf galaxy
data to show that    the stellar masses  and metallicities   of  dwarf
galaxies follow a fairly tight relation, $Z \propto M_{\star}^{2/5}$, and
have argued that blow-out feedback can naturally explain this trend.
Although similar relations were noted previously \citep[e.g.,][]{brodie91a,zaritsky94a},
we refer to this trend as the ``Dekel-Woo relation'', and use
it to motivate and normalize our prescription for 
galactic supernovae ejecta.  The feedback prescription we present below
in fact reproduces the Dekel-Woo relation, but also
gives rise to and explains the [$\alpha$/Fe] 
abundance pattern trends observed for low-mass dwarf galaxies.  
We note that our prescription for gas and metal ejection 
does not strongly affect our 
conclusions regarding the stellar halo progenitor abundance
pattern nor those of our surviving dIrr example.
  The robustness of the model predictions for these subhalos results because
the dark halos of these systems are relatively massive.
However, ejection 
plays an important role in the star formation and enrichment history of
our lower-mass dSph example, where the typical energy 
of supernovae ejecta is significant compared to the binding energy of the halo.

We can estimate the efficiency of gas and metallicity ejection
from dwarf halos, $w$,  by taking the ratio of
specific energy of gas that
escapes the central dwarf galaxy as a wind, $\mathcal{E}_{\rm w}$,
to the specific binding energy of the halo:
\begin{eqnarray}
w \approx \frac{\mathcal{E}_{\rm w}}{\frac{1}{2} \vcirc^2}     \approx      \frac{\epsilon          \eta       E_{\mathrm{SN}}
N_{\mathrm{SN}}}{\frac{1}{2}v_{\mathrm{circ}}^{2}},
\end{eqnarray}
\noindent
where $v_{\mathrm{circ}}$ is  the   circular velocity of   the halo
in question.  
The energy of the gas that leaves the central galaxy
will depend on $N_{\mathrm{SN}}$,  the number of  supernovae  
per unit mass in the stellar
population and $E_{\mathrm{SN}}$, the energy  per supernovae.
The factor $\eta$ is
the fraction of supernovae energy converted into
kinetic energy, and $\epsilon$ is the fraction  of 
the input kinetic energy that is retained
by the gas that actually escapes  the galaxy as  a wind. 
The kinetic energy fraction $\eta$ 
will depend on the speed at which supernovae bubbles
overlap compared to the radiative timescale  of the gas in the galaxy,
and therefore on the supernovae rate \citep{dekel86a}. It  typically
will  not exceed a  few  percent, and is   commonly taken to be 5\%  in
cosmological  simulations of galaxy  formation \citep[c.f.][]{abadi03a}.
With this, typical numbers 
give   $w \approx 22  \epsilon  \eta_{_5} v_{50}^{-2}$,
  where $\eta_{_5} \equiv  \eta/0.05$ and $v_{50}  \equiv \vcirc/50$ km s$^{-1}$,
and we have used $N_{\mathrm{SN}}\approx 1.1 \times 10^{-2}
\Msun^{-1}$, which follows from our chosen IMF  and
stellar population modelling.

The fraction of kinetic energy $\epsilon$ retained by gas that escapes
as a wind will  vary between elements.   While H and He will primarily
not be entrained in the wind because of interactions
with the surrounding ambient medium, $\epsilon \ll 1$, metals
will be    preferentially  blown-out of  galaxies   in the metal-rich
supernovae     ejecta, $\epsilon \sim  1$ \citep{mac_low99a}.
Furthermore,  the rate of Type Ia supernovae per stellar population
is quite slow, dragging out over Gyr timescales, while
the Type II supernovae rate is rapid.  We therefore expect that the
bubble overlap time and  kinetic energy
fraction to be slightly lower for Type Ia
compared to Type II ejecta:  $\eta_{\alpha} \lesssim \eta_{Fe}$.
Motivated  by this physical picture, we set the gas blow-out efficiency
of gas, Fe, and $\alpha$-elements in halos of varying circular
velocities to be
\begin{eqnarray}
w_{\rm gas}(v_{\mathrm{circ}}) & = & 0.15 \times   \left(\frac{v_{\mathrm{circ}}}{50
\mathrm{km}\,\,\mathrm{s}^{-1}}\right)^{-2}, \\
w_{\alpha}(v_{\mathrm{circ}}) & = & 14.4 \times   \left(\frac{v_{\mathrm{circ}}}{50
\mathrm{km}\,\,\mathrm{s}^{-1}}\right)^{-2},\\
w_{Fe}(v_{\mathrm{circ}})  & = & 16.8 \times   \left(\frac{v_{\mathrm{circ}}}{50
\mathrm{km}\,\,\mathrm{s}^{-1}}\right)^{-2}.
\end{eqnarray}
\noindent
We have adopted 
the standard kinetic energy fraction  $\eta = 0.05$ for gas 
and Fe, but a slightly lower value,$\eta = 0.035$,  for the 
$\alpha$ elements.  We have used $\epsilon =  0.01$ for the gas and $\epsilon
= 0.8$ for the metals.

\begin{figure}
\figurenum{2}
\epsscale{1}
\plotone{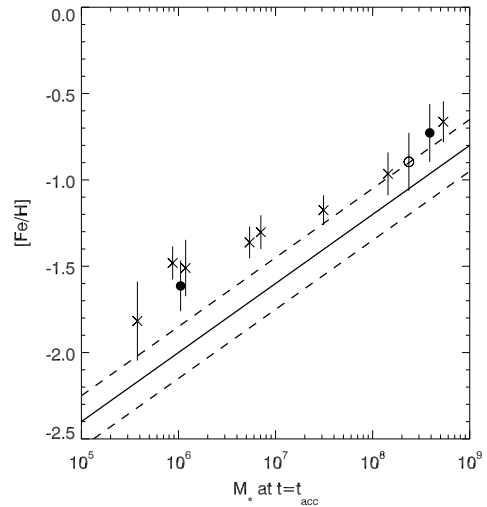}
\caption{\label{fig:dw}
The stellar mass - metallicity  relation for local dwarf galaxies.  We plot the observed trend from
\cite{dekel03a} (solid line), the region containing 70\% of the observed data (dashed lines), and  the calculated relation for a range
in mass
of model   dwarf galaxies (crosses).   The  error  bars  on the  model
calculations delineate the $\sim 1-\sigma$ metallicity range (68\%) of
stars in   each system.   We  use this  calibration  to  constrain the
strength of stellar feedback in our model. 
We also plot our models for surviving dSph and dIrr galaxies (filled circles) and the halo-progenitor model (open circle) for comparison.}
\end{figure}

While  our choices for $w_{Fe}$ and  $w_{\rm gas}$ are of roughly
the expected size, we have chosen their
precise values in order  to reproduce the  observed $M_{\star}-\vcirc$
relation for dwarf  galaxies in the local   group \citep[DW; see][]{bullock05a}
as well as to  approximate the  Dekel-Woo 
mass-metallicity relation.     Figure \ref{fig:dw}  shows   the   stellar-mass
metallicity relation estimated by DW as  the solid line, with
the dashed lines indicating  the $1-\sigma$ scatter in metallicities
about the relation.   The
two solid circles and the open circle show our model results for the
dSph, dIrr, and stellar halo progenitor examples, respectively.
The error bars reflect the 68\% spread in metallicity range for
stars in each model galaxy.  In order to more fully compare
with the observed relation, we have also included 
results for a range of surviving satellite
halos from the models of \cite{bullock05a}.  The mass accretion
histories of these satellites were generated in the same way
discussed \S 2, except now we span a more complete range in  
$\vcirc$.  The model galaxies generally reproduce the observed trend,
with good agreement considering the uncertainty in the data and the
difficulty in assigning precise stellar masses to these observed
systems.
We retain the
stellar feedback calibration that provides this agreement
 for the proceeding calculations.

Note that there is very little freedom in our parameter choices.
Recall that  $\tstar$ and $f_{\gas}$ in our mass accretion model
are already fixed to reproduce the stellar mass of the Galactic halo and
the cold baryon fraction in galaxies.
Once these are given, $w$ is constrained by the
stellar mass-$v_{\rm circ}$ relation, and $w_{\rm Fe}$ by the
approximate stellar mass-metallicity relation.
Finally, the parameter $w_{\alpha}$
is  chosen to reproduce observations of the chemical abundance pattern
of dSph type galaxies
once the above list
of complimentary
observational data is invoked to constrain our other free parameters.
In the next section we show that our choices for the wind
efficiencies indeed reproduce observations of  the chemical abundance pattern
of dSph-type galaxies.
We argue below that
the chemical content of low-mass dwarf galaxies should then show trends
as a function of stellar mass and circular velocity.  Specifically,
we expect that high-mass systems will tend to have near solar
abundance patterns, while low-mass dwarfs will have intermediate
[$\alpha$/Fe] ratios.

\section{Results}
\label{sec:abundances}

\begin{figure*}
\figurenum{3}
\epsscale{1}
\plotone{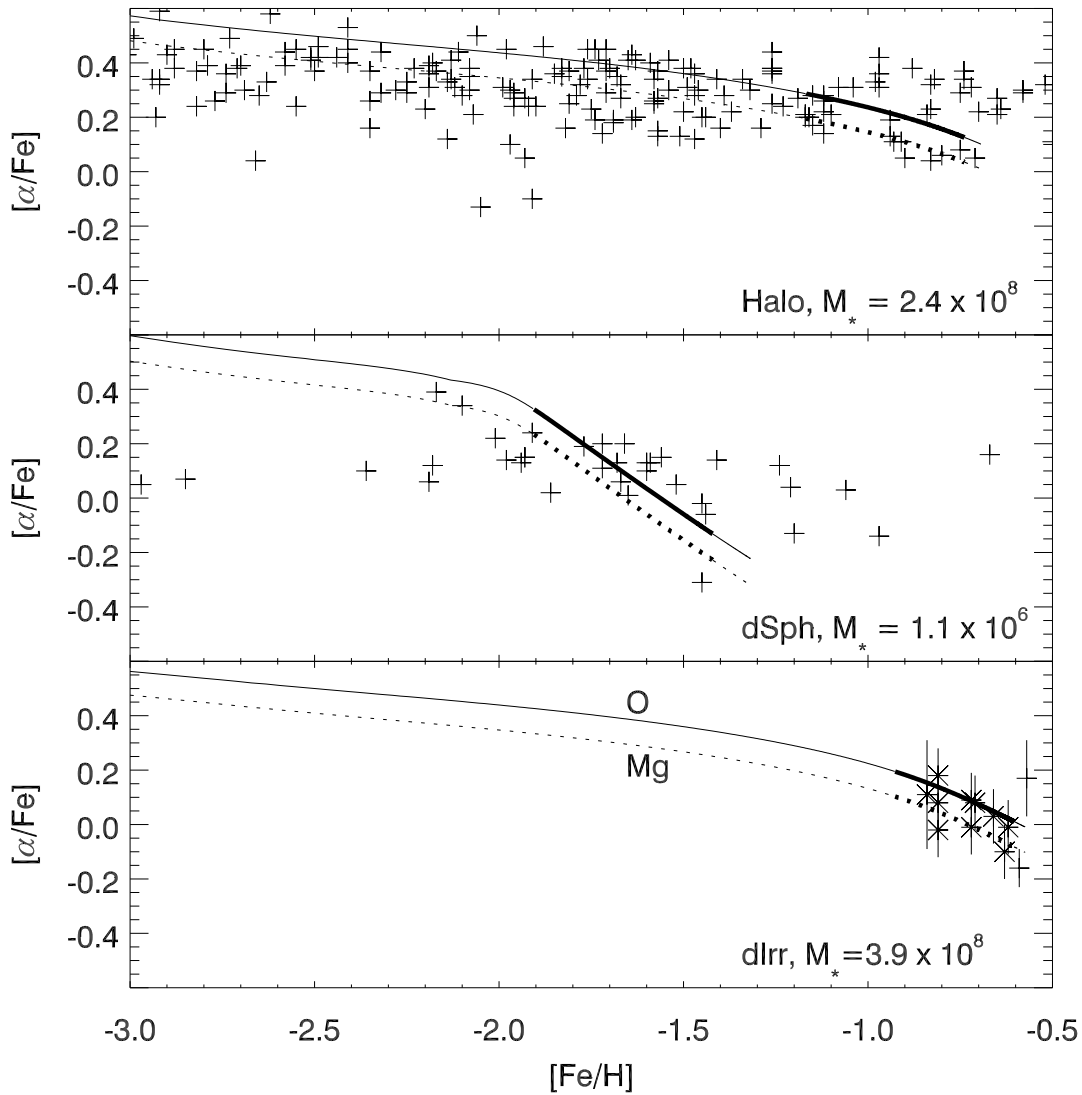}
\caption{\label{fig:afe}
The   [$\alpha$/Fe] vs. [Fe/H]  abundance   of our halo-progenitor  (top
panel), dwarf  spheroidal  (middle  panel), and dIrr-type galaxy
(lower  panel).    {\em{Top:}} The  halo-progenitor model  retains   a super-solar
[$\alpha$/Fe] abundance pattern, resembling the stellar  halo
data  \citep[crosses, from][]{venn04a}. 
{\em Middle:}  The  abundance pattern of  the
dwarf  spheroidal galaxy   is  set by  the  strength  of  the  stellar
feedback, which forces its [$\alpha$/Fe] abundance  below that of halo
stars.
For comparison,
we overplot abundance data of similarly-sized
dSph galaxies
from \cite{venn04a} as crosses. {\em Bottom:}
The majority of the star formation in the
massive  dIrr occurs at  near-solar  values of [$\alpha$/Fe]
and at higher metallicities, similar to  stellar abundances in the SMC
\citep[stars and crosses from][respectively]{venn99a,hill95a}.
Note that we expect extremely metal poor stars in the SMC and LMC to
have enhanced, halo-like [$\alpha$/Fe] ratios. }
\end{figure*}

The three panels in  
Figure \ref{fig:afe} show  the [$\alpha$/Fe] vs. [Fe/H]
tracks computed for our stellar halo progenitor (top),  
dSph example (middle), and
dIrr example (bottom).
 The tracks
  correspond to the  star formation  histories
presented in Figure \ref{fig:sfr}.  In all cases, solid lines show
[O/Fe] and dotted lines show [Mg/Fe].  The final stellar masses
(at accretion) for the systems are indicated in the bottom left of
each panel.  The thick region along each line corresponds to
the abundance region inhabited by the middle
$68\%$ of stars in each system
(the $68 \%$ spread by mass about the median). 

In the upper panel, our massive ($M_0 \simeq 5 \times 10^{10} \Msun$),
early-forming, stellar halo
progenitor 
forms stars quickly until it is accreted  into the Galactic halo $\sim
9$   Gyr  ago.   Its    gas  
enriches  to   [Fe/H]$=-0.7$  and an
$\alpha$-abundance  of [$\alpha$/Fe]=0.1 just before accretion,
while its stars typically have a metallicity of [Fe/H]$\simeq -0.9$ and
[$\alpha$/Fe] $\simeq 0.2$. 
 Upon accretion,  we expect this system
to quickly disrupt, and contribute its stars to the
stellar halo (\S 2).
Over the  predicted chemical evolution
tracks we plot abundance data (crosses) for galactic halo stars from Table 2 of \cite{venn04a}, from which
 we select all stars with Galactic halo membership probabilities greater than
$50\%$
~\footnote{
The data was originally obtained by \cite{fulbright00a,
fulbright02a, gratton88a, gratton91a, gratton94a, hanson98a, ivans03a,
mcwilliam95a, mcwilliam98a, ryan96a,  stephens02a}.}.
  The  halo-progenitor  dwarf  galaxy  abundance   pattern is
consistent with  the overall abundance  pattern of the  Galactic halo.
Agreement between our model and the data arises because the stellar halo progenitor
is accreted and destroyed
before significant Type I enrichment can occur.
More generally, we expect that 
the super-solar $\alpha$-abundance of halo stars is
inherited naturally from  the star  formation and chemical  enrichment
histories   of these massive, early-accreted dwarf galaxies
\citep{font05a}.  As discussed in \S 2, 
the   $\sim 1.5 \times 10^{9}
M_{\sun}$   stellar mass of   the  halo \citep{carney90a} is
expected to be dominated  by
stars from  disrupted systems of this type 
\citep{bullock05a}.

In the middle panel of Figure \ref{fig:afe}, 
our low-mass ($M_0 \simeq 5 \times 10^8 \Msun$),
dSph example 
forms stars more gradually, and is accreted onto the Milky Way halo
$\sim 5$   Gyr  ago.  Metal enrichment is suppressed in this
system compared to the other two panels because the dSph halo
is more loosely bound and supernovae winds can more easily escape (\S 4.2).
Since the overall ejection rate is higher,
the relative efficiency of
blow-out for Type I and Type II supernovae ejecta is more apparent and
creates the
break in the trajectory in the dSph track.
For comparison, the crosses show $\alpha$-abundance
data compiled by 
\cite{venn04a}  for Local Group dwarf spheroidal galaxies.
The  data   reflects   the  well-established  disparity   between  the
[$\alpha$/Fe]$\sim$0.3 abundance ratios of Galactic stellar halo stars
and the [$\alpha$/Fe]$\sim$0.0     - 0.2 abundance  ratios of   local,
low-mass  dwarf galaxies at  similar metallicities.  It is encouraging
that  our  model  dSph evolves  on  an  $\alpha$-abundance  track that
overlaps with data of observed dwarf spheroidal galaxies, with  its final stellar abundance
peaking at [$\alpha$/Fe]$\simeq$ 0.05 and [Fe/H]$\simeq$ --1.65.
We emphasize again that this agreement stems from supernovae blow-out.
The  differences between the
$\alpha$-abundance pattern for our halo progenitor and our dSph example
primarily results from the relative potential well depths of
their host dark matter halos. 
However, we note that differences between the $\alpha$-abundance pattern of specific dSph galaxies may reflect different efficiencies for blow-out of Type Ia and Type II supernovae ejecta between systems \cite[c.f.][]{lanfranchi04a} than what we adopt above.
In principle, differences in blow-out efficiencies between systems will depend on the complicated hydrodynamical problem of following the overlap of supernovae bubbles and cannot be properly addressed by our simple modelling.
Similarly, our choice of initial mass function affects the number of supernovae per unit mass of stars formed, but such alterations to our modeling can be accomodated through our feedback calibration and similar results can be obtained for e.g. the \cite{salpeter55a} IMF.
We therefore rely on the observed trends of $\alpha$-abundance in dSph galaxies, the Dekel-Woo relation, the total stellar mass of the Galactic halo, and the stellar mass-$v_{\mathrm{circ}}$ relation expected for Local Group dwarfs to constrain our singular choices for the blow-out efficiencies that we utilize in our modelling.

Finally, the  [$\alpha$/Fe] vs [Fe/H] track for our massive
($M_0 \simeq 5 \times 10^{10} \Msun$) dIrr example evolves
in a similar manner to that of our stellar halo progenitor
case, except that it is accreted much later and can continue
evolving for several more billion years.
Because of this longer, slower star formation history,
the effect of Type I supernovae 
is  more  pronounced  than in the   halo-progenitor case.
The gas metallicity enriches to [Fe/H] $= -0.6$
with an $\alpha$-abundance  of roughly solar, while  the inner 68\% of
stars by  mass span a  metallicity  of [Fe/H] $=  -0.95$  to [Fe/H] $=
-0.6$ and range from  [$\alpha$/Fe] = 0.2 to  near solar values.   
Indeed, it has abundance  patterns  typical of
massive satellite   galaxies in the local group,
as illustrated by the over-plotted 
from LMC stars by
\cite{hill95a}    and SMC stars by
\cite{venn99a}.  Note that we predict that the lowest metallicity stars in galaxies like the SMC and LMC should have abundance patterns similar to those in halo stars.
The similarity of $\alpha$-abundance patterns in dwarf galaxies at early times suggests we cannot rule out a scenario in which the Milky Way stellar halo was completely comprised from the destruction of dSph galaxies before they were significantly enriched by Type I supernovae.
However, the statistics of dark matter halo accretion histories \citep[see][]{zentner05a} suggest such a scenario is unlikely in the context of hierarchical formation of the stellar halo.
Furthermore, dynamical modelling of the formation of the stellar halo might rule out such a scenario through differences in the resultant spectrum of phase-space correlations. 

Overall, the differences in the shape of the abundance
pattern tracks between the halo, dSph, and dIrr models are
straightforwardly attributable to the mass-dependent strength
of the feedback implementation.  The stronger winds in the
low-mass dSph model cause the [$\alpha$/Fe] abundance to drop
near [Fe/H] $=-2$ as $\sim$70\% of the metals produced by
star formation have been expelled from the system by this time.
At the same metallicity, the halo progenitor and dIrr models have only 
lost $\sim$7\% and $\sim$9\% of their metals, respectively.  For the dSph
model, Type I SN begin to dominate over Type II SN by this
metallicity and the retention of Fe relative to $\alpha$-elements
correspondingly increases.  The relative strength of the $\alpha$-
and Fe-wind efficiencies will vary between individual dSph galaxies
\citep[see, e.g.,][]{lanfranchi04a} to produce their various abundance
patterns, but the [$\alpha$/Fe] tracks of the more massive halo progenitor
and dIrr systems will be fairly insensitive to these details as their
less efficient winds begin to strongly change their abundance patterns
at characteristically higher metallicities.

\section{Conclusions}
\label{sec:conclusions}

By combining cosmologically-motivated star formation and gas accretion
histories  for dwarf  galaxies with  modelling of  the chemical yields
from stellar  populations, we demonstrate  that the abundance patterns
of  the Galactic halo,  low-mass dwarf  spheroidals, and more massive 
dwarf irregular galaxies  can be  understood within the context of
 cosmologically-motivated models of the stellar
halo and dwarf galaxy formation.  
A straightforward prediction of $\lcdm$ is that
most of the stars in the stellar halo were formed in
dIrr-sized dark matter halos that were accreted at
early times and subsequently destroyed.
 The destroyed, proto-halo 
systems are short-lived, form their stars rapidly,
 and are   enriched   predominately by  Type   II
supernovae.  
They are  disrupted before  the   iron input  from Type   I
supernovae can drive them  to higher metallicities, leaving a chemical
abundance pattern that matches observations of 
low-metallicity, $\alpha-$enriched halo stars.

Longer-lived, dIrr-type galaxies that survive to the present day, form
their stars  more gradually.   These systems undergo   both Type I and
Type II supernovae enrichment.  We  presented a ``typical''  dIrr-type
accretion history,  which enriched to a   moderate metallicity with an
abundance pattern similar to that of the SMC and LMC.  We predict that
lower-metallicity stars in  the SMC  or  LMC will  likely show  similar
abundance patterns to those of halo stars.

Low-mass dSph-type galaxies  are likely systems whose chemical abundance
patterns have been  altered significantly  by  supernovae
winds,  as their  shallow
potential  wells cannot  retain 
outflows.  Adopting simple  
assumptions about the ejection wind efficiency,
we showed that the expected population of surviving
 dwarf satellite galaxies in $\lcdm$ galactic halos
match 
the stellar-mass/metallicity relation observed for
 local-group dwarfs (Dekel \& Woo 2003).
We then
argued that the same  stellar  feedback 
is essential in explaining the 
$\alpha-$element 
abundance patterns in dSph galaxies, and demonstrated
this idea by presenting an explicit, cosmologically-motived
dSph model that follows the abundance pattern of the
observed dwarf spheroidals successfully.
Our expectation is that the  abundance  data  for
 dwarf  spheroidal  galaxies as a function of their stellar
mass can be matched by similar, cosmologically-motivated star formation
histories (Font et al., in preparation). 
We predict  that
 [$\alpha$/Fe] ratios should gradually approach the solar
value in dwarf galaxies of increasing 
stellar mass (or velocity dispersion).

As  discussed,  most of  the stars in   the Galactic  stellar halo are
likely remnants of   early-forming dIrr-type galaxies that  chemically
bear little resemblance to present-day dwarf spheroidals.  At the same
time we expect that a  significant  number of dSph-size galaxies  have
been accreted and destroyed over  the Milky Way's history.  Because of
their shallow potential well   depths, these low-mass   dwarf remnants
should be chemically  distinct from most  of the background population
of the stellar halo.  Fossil evidence of these accretion events may be
more easily detected if chemical signatures  are used to help identify
substructure \citep[e.g.][]{nissen97a, font05a}.
We also note that further constraints on our modeling are supplied
by observations of the age and metallicity distribution of halo stars.
While the modeling presented here demonstrates that the chemical
abundance pattern of typical halo progenitors match the observed 
[$\alpha$/Fe] ratio in the Galactic halo, our model must be extended 
account for the full ensemble of halo progenitors expected from the
time-dependent mass spectrum of accreted substructure predicted by
the $\Lambda$CDM scenario for structure formation.  Such an improvement
to our modeling will be presented in future work \citep{font05a}.

While we adopt a feedback formulation commonly used to model chemical evolution in galaxies, we note other feedback models may be more physically motivated.
For instance, hydrodynamical simulations of galaxy formation may include multiphase descriptions of the interstellar medium \cite[][]{mckee77a} that may more naturally produce self-consistent star formation histories for dwarf galaxies as the microphysics of the ISM are better captured \cite[e.g.][]{springel03a}.
Hydrodynamic simulations can also include the effects of supernovae-driven galactic winds formulated in a manner similar to what we employ, but the detailed calculations of the expansion and cooling of shocks allow a reduction in the net number of free parameters \citep[for instance the need for $\epsilon$ is eliminated, see][]{springel03a}.
As these forms of feedback affect the star formation history and gas content of a given object, they will also affect galactic chemical evolution.
We also note it is possible that some dwarf galaxies with spheroidal components contain intermediate-mass supermassive black holes \citep{barth04a,greene04a}, and hydrodynamical simulations of mergers of galaxies containing supermassive black holes with thermal feedback driven by \cite{bondi52a}-style accretion \citep{springel04a,di_matteo05a} demonstrate that supermassive black holes may have considerable impact on the star formation history and gas content of their host systems.
Feedback from supermassive black holes may then significantly alter the chemical evolution history of galaxies in a manner not considered here.
Due to the complex hydrodynamical and radiative cooling issues posed by supernovae and black hole feedback, we suggest self-consistent hydrodynamical simulations that include the relevant physics described above may be necessary to model more completely the effects of feedback on the chemical evolution of dwarf galaxies and the Galactic stellar halo.
However, to demonstrate the feasibility of a cosmologically-motivated explanation for the abundance of the stellar halo we believe our comparatively simple model is sufficient.
We note that our chemical enrichment modeling could be incorporated into the
calculations of \cite{bullock05a} to form a more complete framework to 
model the detailed properties of the Milky Way stellar halo.
We plan to present this model in a future paper \citep{font05a}.

We close by mentioning that our model does
not capture the full complexity of the available observations, notably
of  heavy elements  like Ba,  Y, and  Eu  \citep{venn04a},  and future
inconsistencies between   the scenario for  the  formation of the halo
presented here  and  observations may  arise.  Future modelling  could
conceivably help draw  stronger connections between  the very detailed
stellar observation and cosmological scenarios for the origin of dwarf
galaxies and the  stellar   halo.
For instance, as full chemical
modeling  becomes incorporated  into cosmological  simulations of
galaxy formation, questions about  the chemical abundance of thick disk
stars and accreted  dwarf systems  \citep[c.f.][]{wyse04a} will find 
more satisfying answers \citep[see e.g.,][]{brook04b}.
We therefore encourage future  attempts to
coordinate  detailed chemical modelling with cosmological calculations
to increase the interface between cosmology and galactic astronomy.

\acknowledgements
The authors acknowledge fruitful discussions with K. Venn, M. Shetrone, 
Q. Y. Gong, A. Zentner, and R. Wyse.
This work was supported in part by NSF grants ACI
96-19019, AST 00-71019, AST 02-06299, and AST 03-07690, and NASA ATP
grants NAG5-12140, NAG5-13292, and NAG5-13381.
K.V.J. and A.F.'s contributions were supported through NASA grant NAG5-9064
and NSF CAREER award AST-0133617.

\bibliographystyle{apj}

\end{document}